\documentclass[iop, twocolumn]{aastex63}
\usepackage{verbatim,subfigure}
\hypersetup{citecolor=blue,urlcolor=red}
\shorttitle{Black widow pulsar}
\shortauthors{Wang et al.}
\usepackage{ulem}

\begin{document}

\title{Change of rotation measure during eclipse of a black widow PSR J2051$-$0827}

\correspondingauthor{S.Q. Wang, J.B. Wang, N. Wang}
\email{wangshuangqiang@xao.ac.cn, 1983wangjingbo@163.com, na.wang@xao.ac.cn}

\author{S.Q. Wang}
\affiliation{Xinjiang Astronomical Observatory, Chinese Academy of Sciences, Urumqi, Xinjiang 830011, People's Republic of China}
\affiliation{Key Laboratory of Radio Astronomy, Chinese Academy of Sciences, Urumqi, Xinjiang, 830011, People's Republic of China}
\affiliation{Xinjiang Key Laboratory of Radio Astrophysics, Urumqi, Xinjiang, 830011, People's Republic of China}

\author{J.B. Wang}
\affiliation{Institute of Optoelectronic Technology, Lishui University, Lishui, Zhejiang, 323000, People's Republic of China}

\author{D.Z. Li}
\affiliation{Cahill Center for Astronomy and Astrophysics, California Institute of Technology, 1216 E California Boulevard, Pasadena, CA 91125, USA}

\author{J.M. Yao}
\affiliation{Xinjiang Astronomical Observatory, Chinese Academy of Sciences, Urumqi, Xinjiang 830011, People's Republic of China}
\affiliation{Key Laboratory of Radio Astronomy, Chinese Academy of Sciences, Urumqi, Xinjiang, 830011, People's Republic of China}
\affiliation{Xinjiang Key Laboratory of Radio Astrophysics, Urumqi, Xinjiang, 830011, People's Republic of China}

\author{R.N. Manchester}x
\affiliation{CSIRO Space and Astronomy, PO Box 76, Epping, NSW 1710, Australia}

\author{G. Hobbs}
\affiliation{CSIRO Space and Astronomy, PO Box 76, Epping, NSW 1710, Australia}

\author{N. Wang}
\affiliation{Xinjiang Astronomical Observatory, Chinese Academy of Sciences, Urumqi, Xinjiang 830011, People's Republic of China}
\affiliation{Key Laboratory of Radio Astronomy, Chinese Academy of Sciences, Urumqi, Xinjiang, 830011, People's Republic of China}
\affiliation{Xinjiang Key Laboratory of Radio Astrophysics, Urumqi, Xinjiang, 830011, People's Republic of China}

\author{S. Dai}
\affiliation{School of Science, Western Sydney University, Locked Bag 1797, Penrith South DC, NSW 2751, Australia}

\author{H. Xu}
\affiliation{National Astronomical Observatories, Chinese Academy of Sciences, Beijing 100101, People's Republic of China}

\author{R. Luo}
\affiliation{CSIRO Space and Astronomy, PO Box 76, Epping, NSW 1710, Australia}
\affiliation{Department of Astronomy, School of Physics and Materials Science, Guangzhou University, Guangzhou 510006, China}

\author{Y. Feng}
\affiliation{Zhejiang Lab, Hangzhou, Zhejiang 311121, People's Republic of China}

\author{W.Y. Wang}
\affiliation{Kavli Institute for Astronomy and Astrophysics, Peking University, Beijing 100871, People's Republic of China}

\author{D. Li}
\affiliation{National Astronomical Observatories, Chinese Academy of Sciences, Beijing 100101, People's Republic of China}
\affiliation{NAOC-UKZN Computational Astrophysics Centre, University of KwaZulu-Natal, Durban 4000, South Africa}

\author{Y.W. Yu}
\affiliation{Institute of Astrophysics, Central China Normal University, Wuhan 430079, People's Republic of China}

\author{Z.X. Du}
\affiliation{Institute of Astrophysics, Central China Normal University, Wuhan 430079, People's Republic of China}

\author{C.H. Niu}
\affiliation{National Astronomical Observatories, Chinese Academy of Sciences, Beijing 100101, People's Republic of China}

\author{S.B. Zhang}
\affiliation{Purple Mountain Observatory, Chinese Academy of Sciences, Nanjing 210008, People's Republic of China}

\author{C.M. Zhang}
\affiliation{National Astronomical Observatories, Chinese Academy of Sciences, Beijing 100101, People's Republic of China}

\begin{abstract}

Black widows are millisecond pulsars ablating their companions. The material blown from the companion blocks the radio emission, resulting in radio eclipses. The properties of the eclipse medium are poorly understood. Here, we present direct evidence of the existence of magnetic fields in the eclipse medium of the black widow PSR J2051$-$0827 using observations made with the Five-hundred-meter Aperture Spherical radio Telescope (FAST). We detect a regular decrease in rotation measure (RM) in the egress of eclipse, changing from $60\,\rm rad\,m^{-2}$ to $-28.7\,\rm rad\,m^{-2}$. The RM gradually changes back to normal when the line-of-sight moves away from the eclipse. The estimated line-of-sight magnetic field strength in the eclipse medium is $\sim 0.1$\,G. The RM reversal could be caused by a change of the magnetic field strength along the line of sight due to binary orbital motion. The RM reversal phenomenon has also been observed in some repeating fast radio bursts (FRBs), and the study of spider pulsars may provide additional information about the origin of FRBs. 

\end{abstract}


\section{INTRODUCTION}

Spider pulsars are millisecond pulsars (MSPs) with low-mass companions in short-period orbits~\citep{Fruchter1988,Roberts2013}. They are descendants of low-mass X-ray binaries after accretion on to the pulsar has terminated~\citep{Archibald2009, Stappers2014}. 
Redbacks (RBs) and  black widows (BWs) are  two types of spider pulsars with different companion masses. The companion mass of a RB is in the range of $\sim$0.2$-$0.4 $M_{\odot}$, while that of a BW is much smaller with companion masses $\sim$0.01$-$0.05 $M_{\odot}$~\citep{Roberts2013}. 
In spider pulsars, the pulsar wind and electromagnetic emission ablate and may destroy the companion, leading to the observed isolated MSPs~\citep{Fruchter1988}. 
The material blown from the companion blocks the pulsar radio emission, resulting in radio eclipses. 
Spider pulsars offer valuable opportunities to investigate the characteristics of the companion stars under intense irradiation. 

The study of the eclipse medium is important to understand eclipse mechanisms. 
The radio eclipses of spider pulsars are frequency dependent, with longer duration at lower frequency~\citep{Stappers2001a, Polzin2019}. 
The eclipse mechanism in different systems may be different ~\citep{Thompson1994}. 
By analysis the eclipses of PSR J2051$-$0827 over a wide frequency range of 234$-$1660\,MHz, \citet{Stappers2001a} suggested that pulse scattering, pulse smearing at lower frequencies or cyclotron damping may be possible eclipse mechanisms. 
\citet{Polzin2019} has studied PSR J2051$-$0827 at radio frequencies ranging form 110 to 4032\,MHz, and reached the same conclusions as~\citet{Stappers2001a}. 
By modeling the broadband radio spectrum in the optically thick to thin transition regime of PSR J1544+4937, \citet{Kansabanik2021} found that the observed frequency-dependent eclipses can be well explained by synchrotron absorption by relativistic electrons. \citet{Kumari2023} has revealed a secular variation of the orbital period for PSR J1544+4937.


Polarization analyses of spider pulsars show that there are significant magnetic fields in the eclipse medium, providing a unique opportunity to study the eclipse mechanism of cyclotron damping. 
\citet{Crowter2020} measured RM changes near the eclipse boundary of PSR J2256$-$1024, and constrained the line-of-sight magnetic field to be $\sim 1.11$\, mG.
\citet{Li2019} constrained the line-of-sight magnetic field strength near the eclipse boundary of PSR B1957+20 to be $< 0.02\,$G through plasma lensing.
\citet{Li2022} found evidence of Faraday conversion and attenuation in PSR B1744$-$24A, and estimated the magnetic field to be $\sim 100$\,G.
By analyzing the polarization properties of PSR J2051$-$0827,  \citet{Polzin2019} obtained imprecise limits of the line-of-sight magnetic field of $20\pm120$\,G  and the (near-) perpendicular field of $< 0.3$\,G.

Generally, if the eclipse medium has a significant magnetic field, the RM of pulsar at the orbital phase near the eclipse would show regular orbital variations, which have not been detected in spider pulsars yet. 
Motivated by this, we present highly sensitive radio observation of a bright spider pulsar PSR J2051$-$0827 using FAST to study the polarization properties. 
PSR J2051$-$0827 is the second discovered BW with a spin period of 4.5\,ms and an orbital period of 2.38\,hr~\citep{Stappers1996}.
Follow-up observations have revealed that PSR J2051$-$0827 clearly shows radio eclipses at frequencies bellow 1\,GHz, while radio emission is detected throughout the eclipse at $\sim 1.4\,$GHz~\citep{Stappers2001a}.
The orbital inclination of the binary was estimated to be $\sim 40^{\circ}$ through optical observations~\citep{Stappers2001b}, and more recently, \citet{Dhillon2022} provided an estimate of $\sim 55.9^{\circ}$.

The work in this paper is presented as follows:
in Section 2, we describe our observation  and data processing. In Section 3, we present the results.  We discuss and summarize our results in Section 4.

\section{OBSERVATIONS AND DATA PROCESSING}

The observations of PSR J2051$-$0827 was carried out on January 14, 2022, using the central beam of the 19-beam receiver of FAST with a frequency range of 1.05-1.45\,GHz~\citep{Jiang2019}. 
The data was recorded in search mode PSRFITS format with four polarizations, 8-bit, 1024 frequency channels, and a sampling interval of 16.384\,$\mu$s. 
The polarization response was calibrated using an observation in which a pulsed noise source was injected into the feed before the pulsar observation.  
We used the {\sc dspsr}~\citep{Straten2011} and {\sc psrchive}~\citep{Hotan2004} program to analyze the data set. 
The calibration files were folded at the calibration pulse period of 0.2\,s. 
The data was folded according to the timing ephemeris with a sub-integration time of 10\,s using the {\sc dspsr} program. Then, the data was calibrated using the {\sc psrchive} program {\sc pac} to transform the polarization products into Stokes parameters. More details of the calibration of the FAST data can be found in~\cite{Xu2021}.  
The rotation measure (RM) was measured using the {\sc rmfit} program.

Dispersion measure (DM) is measured using the {\sc tempo2} software package~\citep{Hobbs2006}. 
The data was scrunched to 4 equal frequency channels with a bandwidth of 100\,MHz. 
A noise-free template was formed by fitting the integrated profile of the entire out-of-eclipse observation, and times of arrival (ToAs) were formed by cross-correlating pulse profiles with the standard template.
The DM of each sub-integration was then fitted.

\section{RESULTS}

\begin{figure}
\centering
\includegraphics[width=80mm]{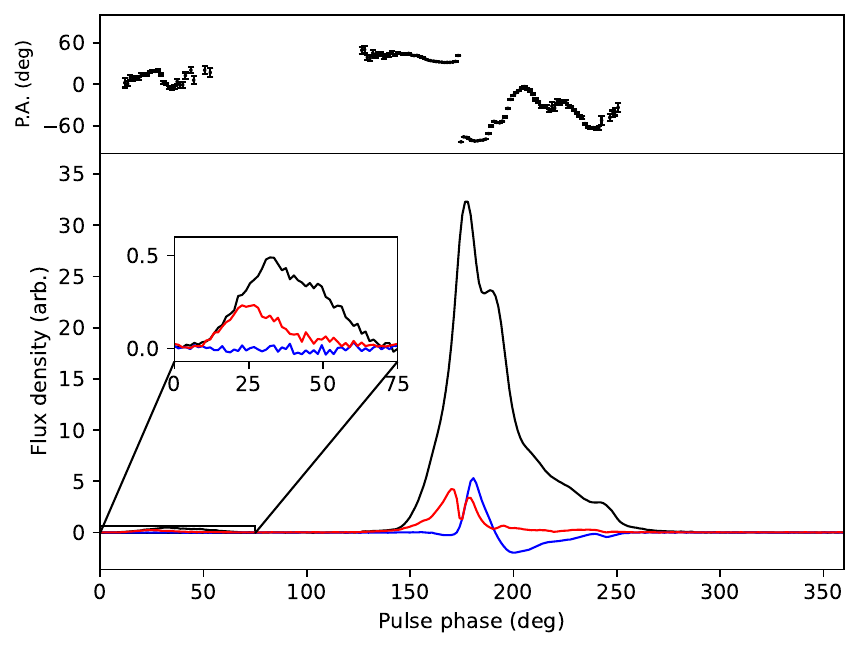}
\caption{Polarization profiles for PSR J2051$-$0827 at 1250\,MHz. The black, red, and blue lines are the total intensity, linear polarized intensity, and circular polarized intensity, respectively. The position angles (black dots) and corresponding uncertainties (black bars) of the linear polarized emission are shown as a function of pulse phase. The inset shows an expanded view of the interpulse. }
\label{prof}
\end{figure}

\begin{figure*}
\centering
\includegraphics[width=160mm]{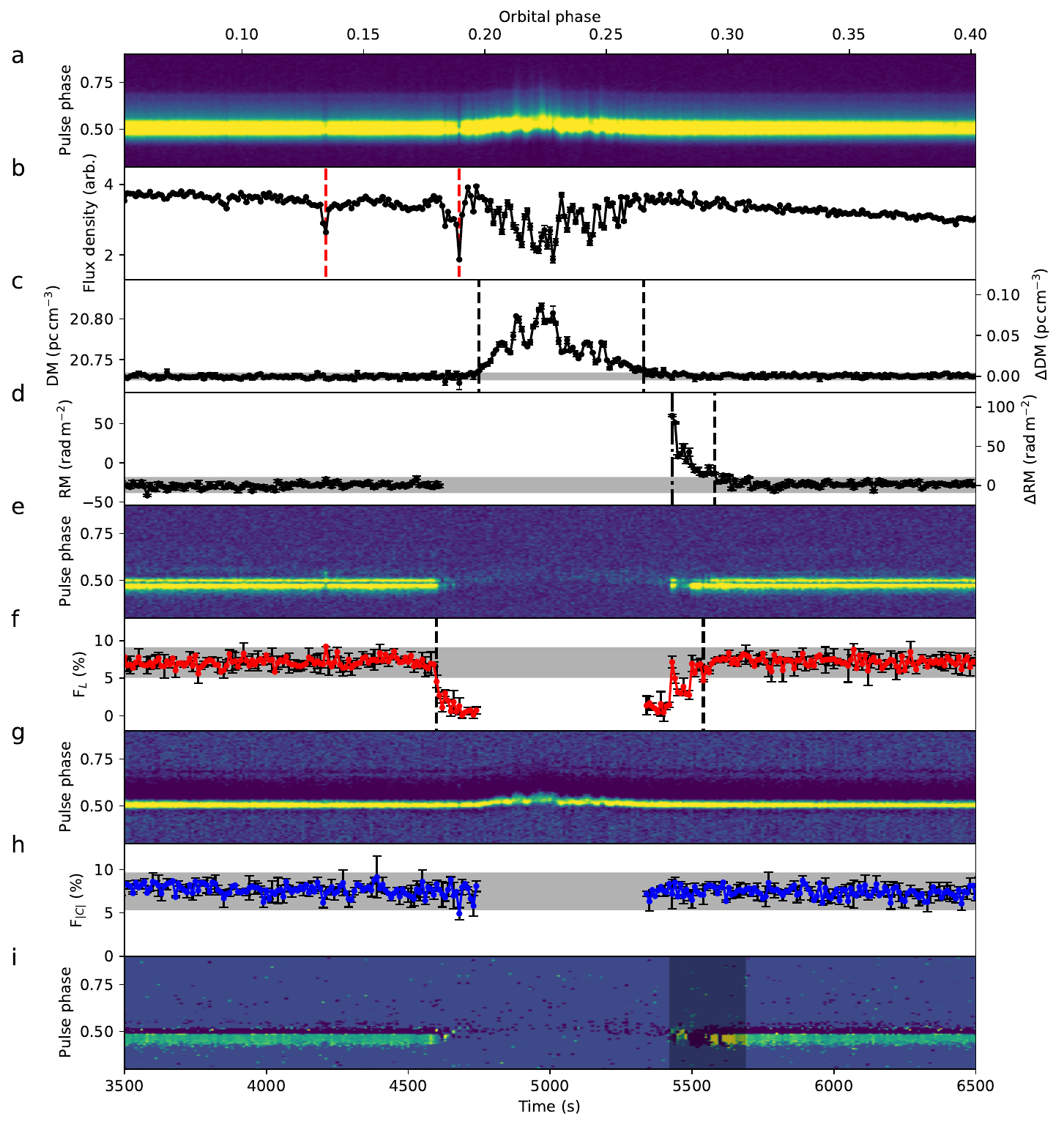}
\caption{The intensity (panel (a)), flux density (panel (b)), DM and $\Delta$DM (panel (c)), RM and $\Delta$RM (panel (d)), the intensity and fraction of linear (panels (e) and (f)) and circular polarizations (panels (g) and (h)), and the PA swings (panel (i)) with a sub-integration of 10\,s of PSR J2051$-$0827 versus orbital phase. 
The filled grey area in the panels (c), (d), (f) and (h) identifies the region of 3$\sigma$. 
The vertical dashed lines the panel (b) label the time where the flux density decreases because of RFI. 
The vertical black dashed line in panels (c), (d) and (f) are the orbital phase where the corresponding parameters show variations larger than $3\sigma$ compared to that of the out-of-eclipse.
Note that we do not show the ${\rm F_{L}}$ and ${\rm F_{V}}$ of sub-integration whose DM variation greater than $3\sigma$. 
The dash-dotted line in the panel (d) labels the orbital phase of maximum RM variation, and the filled grey area in the panel (i) labels the orbital phase where PA swings shift.
The filled grey area in the each panel identifies the region of 3$\sigma$ which is calculated by fitting the distribution of corresponding parameter of sub-integration in out-of-eclipse phase using the Gaussian function. 
}
\label{pola}
\end{figure*}

\begin{figure}
\centering
\includegraphics[width=80mm]{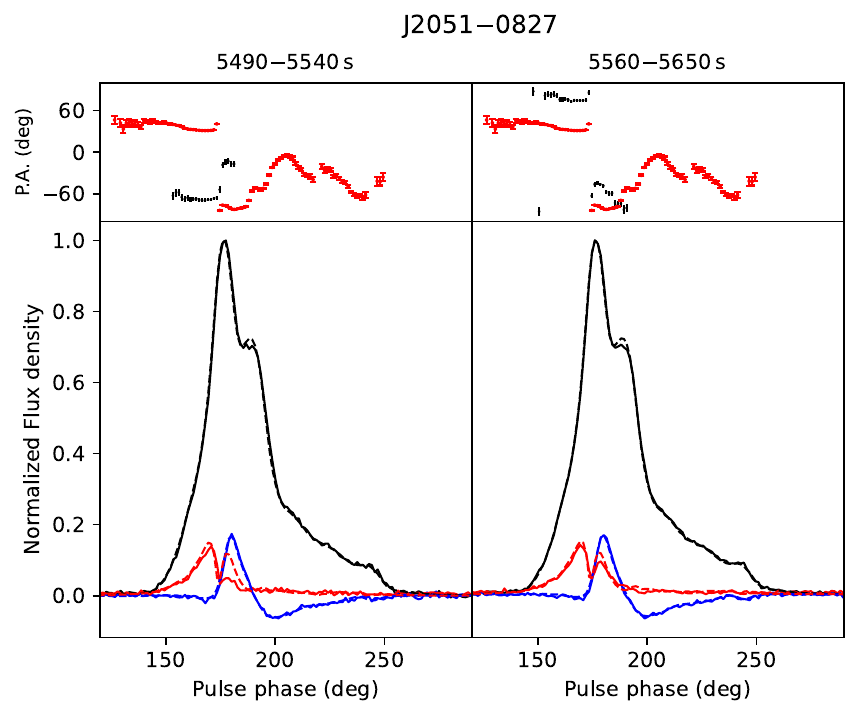}
\caption{Polarization shift profiles between 5490$-$5540\,s (left panel) and 5560$-$5650\,s  (right panel) for PSR J2051$-$0827. The black, red, and blue solid lines are for the total intensity, linear polarized intensity, and circular polarized intensity, respectively. The position angles (black dots) and corresponding uncertainties (black bars) of the linear polarized emission are shown as a function of pulse phase. 
For comparison, the out-of-eclipse polarization profiles are shown in dashed lines, and the corresponding PA swings are shown in red dots.}
\label{pa}
\end{figure}

Figure~\ref{prof} shows the average out-of-eclipse polarization profile of PSR J2051$-$0827 (more details of the eclipse are shown in the following paragraph).
For the first time, a weak interpulse is detected with a flux density only 5\% of the mean pulse.
The measured rotation measure (RM) is $-28.74\pm0.12\,{\rm rad\,m^{-2}}$.
Using the software package IONFR, we obtained the ionosphere contribution to the RM of $2.87\pm0.05\,{\rm rad\,m^{-2}}$.
The corrected RM$_{\rm ISM}$ is $-31.61\pm0.17\,{\rm rad\,m^{-2}}$, which is basically consistent with LOFAR's observations at 149\,MHz~\citep{Polzin2019}. 
The fractional linear ($F_{\rm L}$) and absolute circular ($F_{\rm \vert C\vert}$) polarizations of the main pulse are 8.9\% and 9.3\%, respectively, while those of the interpulse are 36.7\% and 5.7\%, respectively. 
Note that our results of polarization profile agrees with~\citet{Polzin2019}.
Similar to most MSPs, the position angle (PA) swings of PSR J2051$-$0827 exhibits complicated variations. 
There are sense reversals of the circular polarization under the profile of the main pulse. 
A jump of 125\,deg in the PA swings at the pulse phase of 174\,deg is detected, which is different from 90\,deg. Our results suggest that the jump of OPM is frequency-dependent (e.g., \citet{Navarro1997}). 
However, the bandwidth of FAST is only 500\,MHz, further observations with wideband will provide more details of the frequency-dependent OPM of PSR J2051-0827.

Figure \ref{pola} shows the intensity (panel (a)), flux density (panel (b)), DM (panel (c)), RM (panel (d)), the intensity and fraction of linear (panels (e) and (f)) and circular polarizations (panels (g) and (h)), and the PA swings (panel (i)) of PSR J2051$-$0827 with a sub-integration of 10\,s versus orbital phase. 
The pulse intensity becomes weaker, variable dispersion is seen through the shift of the pulse profile during the eclipse, and the radio emission is detected throughout the eclipse at 1250\,MHz, {which agrees with the former results (e.g., ~\citealt{Stappers2001a, Polzin2019}).}

During the eclipse, the radio emission propagates through the ionized eclipse medium, which could result in an increase in DM. 
In our observation, the DM shows a significant increase between the orbital phase of $0.198$ and $0.265$, with a duration of $\sim$580\,s, about 7\% of the orbital period  (the vertical dashed lines in the panel (c) of Figure~\ref{pola}).
The maximum DM during the eclipse is 20.815$\pm$0.003\,$\rm cm^{-3}\,pc$ with a $\Delta \rm DM_{\rm max}$ of 0.085\,$\rm cm^{-3}\,pc$ compared to that of the average out-of-eclipse profile of 20.7296$\pm$0.0003\,$\rm cm^{-3}\,pc$. 
We detect a regular decrease in RM during the eclipse egress, changing from $60\,\rm rad\,m^{-2}$ to $-28.7\,\rm rad\,m^{-2}$, in which the RM gradually change back to normal when the line-of-slight moves away from the eclipse medium (the panel (d) of Figure~\ref{pola}). 
The PA swings of linear polarization also shift during the eclipse egress (the filled grey area in the panel (i) of Figure~\ref{pola}, see also Figure~\ref{pa}). 
{ Note that we used RMFIT to measure the RM values of the two profiles at the orbital phase where the PA shift is detected (Figure~\ref{pa}), and obtained that the $\rm RM_{\rm ISM}$ of them are $-14 \pm 1$\,$\rm rad\,m^{-2}$ and $-19.0 \pm 0.7$\,$\rm rad\,m^{-2}$, respectively. The profiles are then corrected, and are shown in Figure~\ref{pa}. }
{Compared to the average polarization profile in the out-of-eclipse region, the profile at the orbital phase where the PA shift is detected is almost unchanged, as well as the circular polarization, but the linear polarization is changed.}
However, we cannot measure the level of PA swings shift because of the complicate PA swings variations of PSR J2051$-$0827.
Our result provides direct evidence of the existence of magnetic field in the eclipse medium in spider pulsars.
Although RM variations have been detected in some spider pulsars~\citep{Polzin2019,Crowter2020, Li2019,Li2022}, no regular decrease/increase in RM near eclipse has been detected yet. 
We believe that we observed a regular decrease/increase in RM in spider pulsars for the first time, which provides a strong evidence of the existence of magnetic field in the eclipse medium.
The RM reversal is possibly caused by changes of the magnetic field strength along the line of sight due to orbital motion, and the direction of the line-of-sight magnetic field of eclipse medium is opposite to that of the medium between the binary system and the earth. 
No significant variations in RM is detected during ingress.
In spider pulsars, the eclipse material is shocked by the pulsar wind, leaving a cometary-like tail of magnetic field of material. As in the ingress, the RM is changing faster than the egress and hence more susceptible to depolarization after summing over pulses.

During the  egress of the eclipse, the RM variation ($\Delta$RM) decreases within 3$\sigma$ at the orbital phase of 0.295 (the vertical dashed line in the panel (d) of Figure~\ref{pola}), and shows maximum variation at the orbital phase of 0.277 (the vertical dash-doted line in panel (d) of Figure~\ref{pola}).
{The maximum $\rm RM_{\rm ISM}$ during eclipse is about $57$\,$\rm rad\,m^{-2}$ with a $\Delta  \rm RM_{\rm max}$ of 89\,$\rm rad\,m^{-2}$ compared to that of the out-of-eclipse.}
The $\Delta \rm DM$ at the orbital phase of $\Delta  \rm RM_{\rm max}$ is 0.001\,$\rm cm^{-3}\,pc$.
The line-of-sight magnetic field strength in the eclipse medium can be estimated by measuring the changes in the Faraday rotation:
\begin{equation}
B_{\|}={\rm 1.23\mu G \, \frac{ \Delta RM}{ \Delta DM}}.
\end{equation}
For PSR J2051$-$0827, we obtain the line-of-sight magnetic field strength of the eclipse medium $B_{\vert \vert} \sim 0.1$\,G which agrees with the former result of $20\pm 120$\,G~\citep{Polzin2019}.
The circular polarization remains unchanged at the orbital phase of $\Delta  \rm RM_{\rm max}$ (the panels (g) and (h) of Figure~\ref{pola}), which is expect for the magnetic filed $< \sim 100$\,G~\citep{Li2019, Li2022}.

The depolarization phenomenon is observed during eclipse of PSR J2051$-$0827. 
During the eclipse, the RM between the orbital phase 0.182 and 0.276 cannot be determined (the panel (d) of Figure~\ref{pola}), which may be due to the rapid RM fluctuations caused by the turbulence in eclipse medium~\citep{You2018}. 
At eclipse boundary, the linear polarization begins/ends to decrease at the orbital phase of 0.180 and 0.290 (the vertical dashed lines in panel (f) of Figure~\ref{pola}), while both the circular polarization and DM remain unchanged. 
This suggests that there may be a small-scale structure in electron column density or magnetic field at the boundary of the eclipse medium. 
The decreases of linear polarization could be caused by large RM variations in a sub-integration. 
Assuming the RM variation has a normal distribution with a standard deviation $\sigma_{\rm RM}$, the depolarization due to RM variations could be described as~\citep{You2018}: $L=L_{0}\exp(-2\lambda^4\sigma_{\rm RM}^2)$, 
where $L_0$ and $L$ are the linear polarization magnitude before and after the integration over fluctuations. 
The $L_{\rm 0} $ of PSR J2051$-$0827 is 8.9\%. 
The $\sigma_{\rm RM}$ of 18\,$\rm rad\,m^{-2}$ could cause  $L $ to be at the level of 1\%.

\section{DISCUSSION AND CONCLUSIONS}

The properties of the eclipse medium is important to understand eclipse mechanisms.  The measurement of the magnetic field of the eclipse medium in spider pulsars provides an opportunity to study the eclipse mechanism of cyclotron damping~\citep{Crowter2020,Polzin2020, Li2022}.
The dependence between the frequency of damped radiation and the magnetic field strength in the eclipse medium is described as: 
\begin{equation}
\frac{\nu_{\rm d}}{1\,{\rm GHz}} \approx 2.8\times10^{-3} \frac{B}{\gamma_{\rm p}(1-\cos \theta)},
\end{equation}
with the frequency of radiation that is damped $\nu_{\rm d}$, the Lorentz factor of particles in the eclipse medium $\gamma_{\rm p}$, the angle between the wave propagation and the magnetic field $\theta$. 
Assuming $\gamma_{\rm p}(1-\cos \theta)\sim$ 1, we find that observations at 1.25\,GHz could be affected for $B > 450\,$G. 
However, our estimated value of $B_{\vert \vert}$ for the eclipse medium of PSR J2051$-$0827 is approximately $0.1$\,G, and \citet{Polzin2019} obtained an imprecise limit of PSR J2051$-$0827 of $20\pm 120$\,G, both of which are too low to significantly impact observations at 1.25\,GHz. 
The magnetic field of the eclipse medium has been measured in some spider pulsars, e.g. PSR J2256$-$1024 with $B_{\vert \vert} \sim 1.11$\,mG~\citep{Crowter2020}, PSR B1957+20 with $B_{\vert \vert} < 0.02$\,G~\citep{Li2019}, PSR B1744$-$24A with $B \sim 100$\,G~\citep{Li2022}.  
Considering the magnetic field levels in the eclipse medium of spider pulsars, cyclotron damping may not be the primary eclipse mechanism at L-band.
{However, different systems may have different eclipse mechanisms. \citet{Kansabanik2021} found that the observed frequency dependent eclipses for PSR J1544+4937 can be well explained by synchrotron absorption by relativistic electrons.
\citet{Bai2022} presented measurements of pulse scattering near the eclipse of PSR B1957+20 using FAST and found that scattering is one of the main eclipse mechanisms at L-band. 

A clear frequency dependence of the eclipse duration has been detected in many spider pulsars, which could be modeled by a power-law exponent with an index in the range of $0.19-0.68$  (e,g., \citealt{Nice1990,Polzin2020}). 
However, for PSR J1816+4510 and PSR B1957+20, a single power law is insufficient to model the eclipse duration across a wideband~\citep{Polzin2020}. This deviation from a power law could be caused by a tenuous swept-back tail of material~\citep{Fruchter1990}. It is possible that the properties of the tail material differ from the main bulk of the medium, and the eclipse mechanisms may vary at different frequencies. 
The ultra-wideband receiver (UWL) on the Parkes radio telescope, which provides continuous frequency coverage from 704 to 4032 MHz~\citep{Hobbs2020}, is an idea tool for studying the frequency dependence of the eclipse duration in spider pulsars. Further observation using the UWL will provide additional constraints on the eclipse mechanism. 
}

The companion of PSR J2051$-$0827 is is identified as a brown dwarf~\citep{Dhillon2022}. 
Assuming that the companion has a dipolar field, we can describe the magnetic field strength at its surface as: $B_{\rm BD}=B_{\rm e}(R_{\rm e}/R_{\rm BD})^{3}$, where $R_{\rm BD}$ and $R_{\rm e}$ are the radius of the companion and eclipse medium, respectively. The radius of a brown dwarf is generally falls within the range of $0.64-1.13\,M_{\rm J}$~\citep{Sorahana2013}.
For PSR J2051$-$0827, taking $B_{e} \sim 0.1$\,G and $R_{\rm e} = 0.65\,R_{\odot}$, the magnetic field strength at the surface of the companion is estimated to be in the range of $\sim 20-100$\,G. By detecting the circularly polarized pulses, \citet{Kao2018} measured the localized magnetic fields of brown dwarfs to be about several kG. 
However, their observations are made at high frequencies (C/X-band), which are more favorable for detecting  strong magnetic field brown dwarfs.

The environments of spider pulsars exhibit some similarities with some FRBs. The depolarization phenomenon and Faraday conversion have been seen in both of spider pulsars and some FRBs~\citep{Xu2021,Feng2022,Li2022}. 
Therefore, the study of spider pulsars could potentially provide additional insights into the origin of FRBs.
Neutron stars in binary systems have been proposed as possible origins of FRBs~\citep{Zhang2017,Margalit2018,Lyutikov2020}. Radio emissions from the neutron stars are expected to propagate through surrounding dense and ionized medium in the binary system. 
The large and variable RM observed in FRBs  indicates that FRBs are in an extreme and dynamic magneto-ionic environment~\citep{Anna-Thomas2023,Xu2021, Feng2022, Michilli2018,Luo2020, Hilmarsson2021}. 
Recently, \citet{Anna-Thomas2023} reported the reversal of RM of FRB 20190520B, which changed from $\sim 10000\,\rm rad\,m^{-2} $ to $\sim  -16000\,\rm rad\,m^{-2}$. 
A similar RM reversal has been observed near the eclipse of PSR J2051$-$0827. 
During the egress of the eclipse, the RM of PSR J2051$-$0827 changed from $60\,\rm rad\,m^{-2}$ to $-28.7\,\rm rad\,m^{-2}$, gradually returning to normal as the line-of-sight moved away from the eclipse medium. 
This RM reversal could be attributed to changes in the magnetic field strength along the line of sight due to the binary orbital motion.
Noted that the direction of the line-of-sight magnetic field in the eclipse medium is opposite to that of the medium between the binary system and Earth. This mechanism may also be applicable to FRB 20190520B.

Assuming that the RM variation of FRB is caused by changes in the surrounding magneto-ionic cold plasma,  the RM variation is described as:
\begin{equation}
{\rm \Delta RM}=0.81\, {\rm rad\,m^{-2}}  \int_{0}^{l} \frac{B_{\|}(l) n_{e} (l)}{(1+z(l))^2}dl,
\end{equation}
where $l$ is the geometric length scale of the surrounding medium, $B_{\|}$ is the line-of-sight magnetic field strength in microgausss, $z$ is the redshift of the source.  
For FRB 20190520B, with $\Delta {\rm RM} \sim 10^4\,{\rm rad\,m^{-2}} $~\citep{Anna-Thomas2023}, the estimated $B_{\|} n_{e}  \sim 100\, {\rm cm^{-3}}\,$G\,pc. 
In the case of the surrounding medium of PSR J2051$-$0827, where we measured $B_{\|} \sim$ 0.1\,G and $n_{\rm e}\sim 10^5\, \rm cm^{-3}$ with  $l=0.4\,R_{\odot}$, the estimated value of $B_{\|} n_{e}$ is much lower. 
These results suggest that the surrounding medium of FRBs is more extreme compared to that of spider pulsars. 
Additionally, a strong magnetic field at the Gauss level in the surrounding medium of FRB 20201124A was reported~\citep{Xu2021}. 
In the binary scenario, the observed RM would exhibit periodic evolution for an edge-on system. 
Long-term monitoring of FRBs with RM measurements will provide further constraints on the origins of FRBs.

\section*{Acknowledgments}

This is work is supported by 
the National Natural Science Foundation of China ( No. 12203092, No. 12288102, No. U1938117),
the Natural Science Foundation of Xinjiang Uygur Autonomous Region (No. 2022D01B71, No. 2022D01D85), the Tianchi Doctoral Program and the Special Research Assistant Program of CAS, 
the Major Science and Technology Program of Xinjiang Uygur Autonomous Region (No. 2022A03013-2, No. 2022A03013-3, No. 2022A03013-4), 
the National SKA Program of China (No. 2020SKA0120100), 
the National Key Research and Development Program of China (No. 2022YFC2205201), 
the CAS Project for Young Scientists in Basic Research (No. YSBR-063), the Key Research Project of Zhejiang Laboratory (No. 2021PE0AC03). 
the Zhejiang Provincial Natural Science Foundation of China under Grant (No. LY23A030001), 
This work made use of the data from the Five-hundred-meter Aperture Spherical radio Telescope, which is a Chinese national mega-science facility, operated by National Astronomical Observatories, Chinese Academy of Sciences.

\software{DSPSR \citep{Straten2011}, PSRCHIVE \citep{Hotan2004} and TEMPO2 \citep{Hobbs2006}}

\bibliography{sample63}{}
\bibliographystyle{aasjournal}

\end{document}